\documentclass[aps,twocolumn,floatfix,showpacs]{revtex4}
\usepackage{graphicx}
\usepackage{ae}

\begin{document}

\title{Instabilities of a matter wave in a matter grating}
\author{Giovanni Barontini and Michele Modugno}
\address{LENS \& Dipartimento di Fisica, Universit\`a di Firenze,
Via N. Carrara 1, 50019, Sesto Fiorentino, Italy}
\date{\today}

\begin{abstract}
We investigate the stability of Bloch waves for a Bose-Einstein condensate moving through a periodic lattice created by another condensate modulated by an optical lattice. We show that the coupling of phonon-antiphonon modes of the two species give rise to a very rich structure of the regimes for dynamical instability, with significant differences with respect to the case of a single condensate in an optical lattice. We characterize the relative weight of each condensate in the mixing and discuss an analytic limit that accounts for the bare structure of the instability diagrams.
\end{abstract}

\maketitle

Bose-Einstein condensates (BECs) in optical lattices have proven to be a versatile tool for investigating fundamental issues of quantum mechanics, solid state physics, nonlinear systems and superfluidity \cite{morsch,bloch}. Already in 1D, and in the mean field regime, the dynamics of BECs in a periodic potential presents nontrivial features emerging from the interplay of nonlinearity -- arising from atomic interactions -- and the discrete translational invariance of the lattice. Noticeable features are the formation of solitonic structures \cite{solitons} and the presence of several kinds of instabilities \cite{wuniu,smerzi,menotti, machholm,fallani,desarlo,modugno,barontini,javanainen}. 
In particular, a long investigated issue has been the dynamical instability of nonlinear Bloch waves in the flow of a BEC in a 1D optical lattice. This phenomenon is characterized by the exponential growth of arbitrarily small fluctuations that may lead to a partial or complete destruction of the condensate \cite{wuniu,fallani,modugno} or to the occurrence of a pulsating behavior \cite{barontini,javanainen}, depending on the interaction regime. 
Recently, also the dynamical stability of a two component BECs has been investigated theoretically \cite{hooley, ruostekoski}.

All the studies performed so far, however, have been restricted to the case where the periodic lattice is not perturbed by the condensate flow, since optical lattices cannot support excitations. Different scenarios may occur instead when excitations of the lattice interact with that of the condensate, like electrons and phonons in a crystal lattice. This condition can be realized, for example, by considering a two component BEC system, where one of the two condensates is periodically modulated by an external optical potential, playing the role of a periodic lattice where the other condensate (transparent to the optical potential) flows through.
This configuration can be designed experimentally by suitably tuning over the atomic transitions the electric dipole atom-light interaction that is used to manipulate the atoms. 
In particular, for alkali atoms in case of large detuning and negligible saturation, the intensity of the dipole potential is given by \cite{grimmopt}
\begin{equation}
U_{dip}\propto I\left(\frac{2+Pg_F m_F}{\Delta_{2,F}}+\frac{1-Pg_F m_F}{\Delta_{1,F}} \right),
\label{eqpot}
\end{equation}
where $I$ is the intensity of the laser, $\Delta_{x,F}$ are the detunings from the two fine transitions ($D_1$ e $D_2$) and $P$ is the polarization of the light ($P=0,\pm 1$ for linear and $\sigma^{\pm}$ polarization, respectively). 
For ultra-cold atomic mixtures, since the frequencies of the fine transitions change with the species, the potential produced by the same light acts in different ways with different atomic species, and one can take advantage of this feature by choosing the wavelength of the laser so that the potential in Eq. (\ref{eqpot}) vanishes for one species but not for the other. 

Actually, a double species BEC with tunable interspecies interactions has been recently produced and characterized \cite{thalhammer}. 
Such a system represents indeed a very promising tool for exploring novel configurations of ultracold atomic mixtures in optical lattices \cite{catani}.
\begin{figure}[!t]
\centerline{\includegraphics[width=0.85\columnwidth]{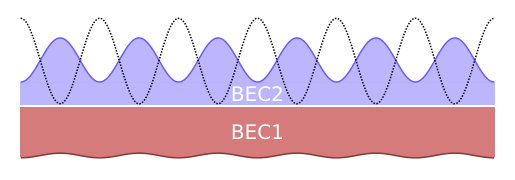}}  
\caption{(Color online) Schematic representation of the system considered: 
in an interacting two component BEC, one of the two condensates (BEC2) is periodically modulated by an external optical lattice (dotted line), and acts as periodic potential through which the other condensate flows (BEC1, that is not affected by the optical potential).}
\label{fig1}
\end{figure}

In this Letter we study the stability of a one-dimensional system composed by two condensates with repulsive interactions, under the effect of an optical lattice to which one of the two species is transparent. In this situation it is possible to investigate the effects of the mutual interaction in the flow of a non-linear matter wave (the condensate that experiences no optical potential, hereinafter BEC1) through a matter grating (the one modulated by the optical lattice, BEC2), as depicted in Fig. \ref{fig1}.
We find that, as far as dynamical instability is concerned, this system exhibits a very different and reacher behavior respect to the usual case of a single BEC moving in an optical lattice. In particular, the conditions for which BEC1 becomes dynamical unstable (BEC2 being at rest in the lattice) resemble the case of instability for a single \textit{attractive} condensate \cite{barontini}, with the addition of novel instability regions due to the interspecies interaction. Such instability regions mix up with those for a single repulsive condensate when also BEC2 is moving with respect to the lattice. We map out the
different regimes for energetic and dynamical instability, characterizing the
growth rates of the unstable modes and the relative weight of each condensate in the mixing. We also discuss an analytic limit for vanishing lattice and vanishing interspecies interaction, that accounts for the bare structure of the instability diagrams in terms of the coupling of phonon-antiphonon modes of the two species.

The above system can be described by two coupled 1D Gross-Pitaevskii equations 
\begin{eqnarray}
\label{2GPE}
i\partial_{t}\Psi_{1}&=&\left[ -\lambda\frac{\partial^2}{\partial x^2}+ g_{11} |\Psi_1|^2+g_{12}|\Psi_2|^2\right]\Psi_1 \\
i\partial_{t}\Psi_2 &=&\left[ -\frac{\partial^2}{\partial x^2}+s\cos^2(x) +g_{21}|\Psi_1|^2+g_{22} |\Psi_2|^2\right]\Psi_2 \nonumber\end{eqnarray}
that are written here in dimensionless form, with lengths in units of the lattice wavevector $k_B=2\pi/\lambda_{opt}$, and the height of the lattice $s$ in units of the recoil energy  $E_R=\hbar^2k_B^2/2m_2$.
The coupling constants can be expressed in terms of the scattering lengths $a_{ij}$ and of the mean densities in each site of the lattice $n_{i}$ as $g_{ij}=2^{2+\delta_{ij}}\pi \sqrt{n_{i}n_{j}} a_{ij}\lambda^{\delta_{i1}\delta_{j1}}[1+\lambda(1-\delta_{ij})]/k_B^2$, 
with $\lambda=m_2/m_1$ (notice that $g_{12}=g_{21}$). With this choice, the wavefunctions $\Psi_{i}$ are normalized to unity over a single period.

To illustrate a specific case, here we consider the case of a $^{87}$Rb -$^{41}$K mixture, with an optical lattice of wavelength $\lambda_{opt}=$ 790 nm and linear polarization. This choice ensures that the condensate of $^{87}$Rb will experience no optical potential, playing therefore the role of BEC1, while the $^{41}$K is subjected to the optical lattice, being BEC2. We have also fixed the mean densities to $10$ particles per lattice site, corresponding to the following values of the intraspecies couplings: $g_{11}\simeq0.021$,  $g_{22}\simeq 0.027$. However, the effects we are going to present can be obtained in a wide range of parameters, with different species or different lattice periodicity.

Equation (\ref{2GPE}) admits stationary solutions in the form of Bloch waves $\Psi_{j}(x)=e^{i(\mu_{j} t-kx)}\phi_j(k,x)$,  where $\phi_j(k,x)$ has the same period of the lattice, the Bloch wave vector $k$ represents the quasimomentum of the condensate (restricted to the first Brillouin zone \cite{wuniu}), and $\mu_{j}$ is the chemical potential of each species.
In the following, for investigating the stability of these Bloch states, we consider three different configurations: \textit{A)} the BEC1 propagates through the BEC2 at rest 
($k_{1}=k$, $k_{2}=0$); \textit{B)} analogous to the former, but with the role of the two condensate exchanged ($k_{1}= 0$, $k_{2}=k$); \textit{C)} both condensate are moving with the same quasimomentum ($k_{1}=k_{2}=k$).

As for the case of a single condensate, two kinds of instability are relevant, namely 
energetic (or Landau) and dynamical instabilities, and can be addressed by considering the stability of the system against small perturbation \cite{wuniu}. For the case of energetic instability the excitations can be decomposed in terms of Bloch waves of quasimomentum $q$ as
\cite{wuniu}
\begin{equation}
\delta\phi_{j}=\frac{1}{\sqrt{n_{j}}}
\left(u_{k,q}^{(j)}(x)e^{iqx}+v_{k,q}^{(j)*}(x)e^{-iqx}\right)
\end{equation}
where the functions $u$ and $v$ have the same spatial periodicity of the lattice. They satisfy the following eigenvalue equation 
\begin{equation}
M(q)X= E(q)X\,,\qquad X=\pmatrix{u^{(1)} \cr v^{(1)}\cr u^{(2)} \cr v^{(2)}}
\end{equation}
obtained by considering the second order contribution to the energy functional \cite{wuniu}.
The matrix $M(q)$ is given by
\begin{equation}
M(q)=\pmatrix{M_{k_{1},q}^{(1)}   & I_{12} \cr
                  I_{21} & M_{k_{2},q}^{(2)} \cr}
\end{equation}
with
\begin{equation}
M_{k,q}^{(i)}=\pmatrix{\mathcal{L}_i(q+k) & g_{ii} \phi_i^2    \cr
 						 g_{ii} \phi_i^{*2} & \mathcal{L}_i(q-k) \cr}
\end{equation}
and
\begin{eqnarray}
\mathcal{L}_j(k)&=&
-\lambda^{2-j}\left(\frac{\partial^2}{\partial x^2}+ik\right)^2+(j-1)s\cos^2(x) \nonumber\\
&&\quad -\mu_j +2g_{j,j}|\phi_j|^2 +g_{j,3-j}|\phi_{3-j}|^2
\end{eqnarray}
\begin{equation}
I_{ij}=g_{ij}\pmatrix{\phi_i \phi_j^* & \phi_i\phi_j \cr
                       \phi_i^*\phi_j^* & \phi_i^*\phi_j \cr}.
\end{equation}
The energetic instability occurs when the condensate flow corresponding to a Bloch wave is no more a local minimum of the energy functional, that is when the $M_k(q)$ matrix has negative eigenvalues \cite{wuniu}.

For the case of dynamical instability we look for time-dependent fluctuation of the form
$$
\delta\phi_{j}=\frac{1}{\sqrt{n_{j}}}
\left(u_{k,q}^{(j)}(x)e^{i(qx-\omega_{kq}t)}+v_{k,q}^{(j)*}(x)e^{-i(qx-\omega_{kq}t)}
\right)
$$
yielding the following Bogoliubov equations \cite{wuniu,hooley,ruostekoski}
\begin{equation}
\Sigma_3 M(q)X=\omega_{kq}X\,,\qquad\Sigma_{3}=
\pmatrix{\sigma_{3} & 0\cr 0& \sigma_{3}}
\end{equation}
where $\sigma_3$ is the third Pauli matrix.
Dynamical instability occurs when some of the eigenfrequencies $\omega_{kq}$ get a nonzero imaginary part, indicating an exponential growth of the corresponding modes, that rapidly drives the system away from the steady state in the dynamical evolution.

\begin{figure*}[!t]
\centerline{\includegraphics[width=0.9\columnwidth]{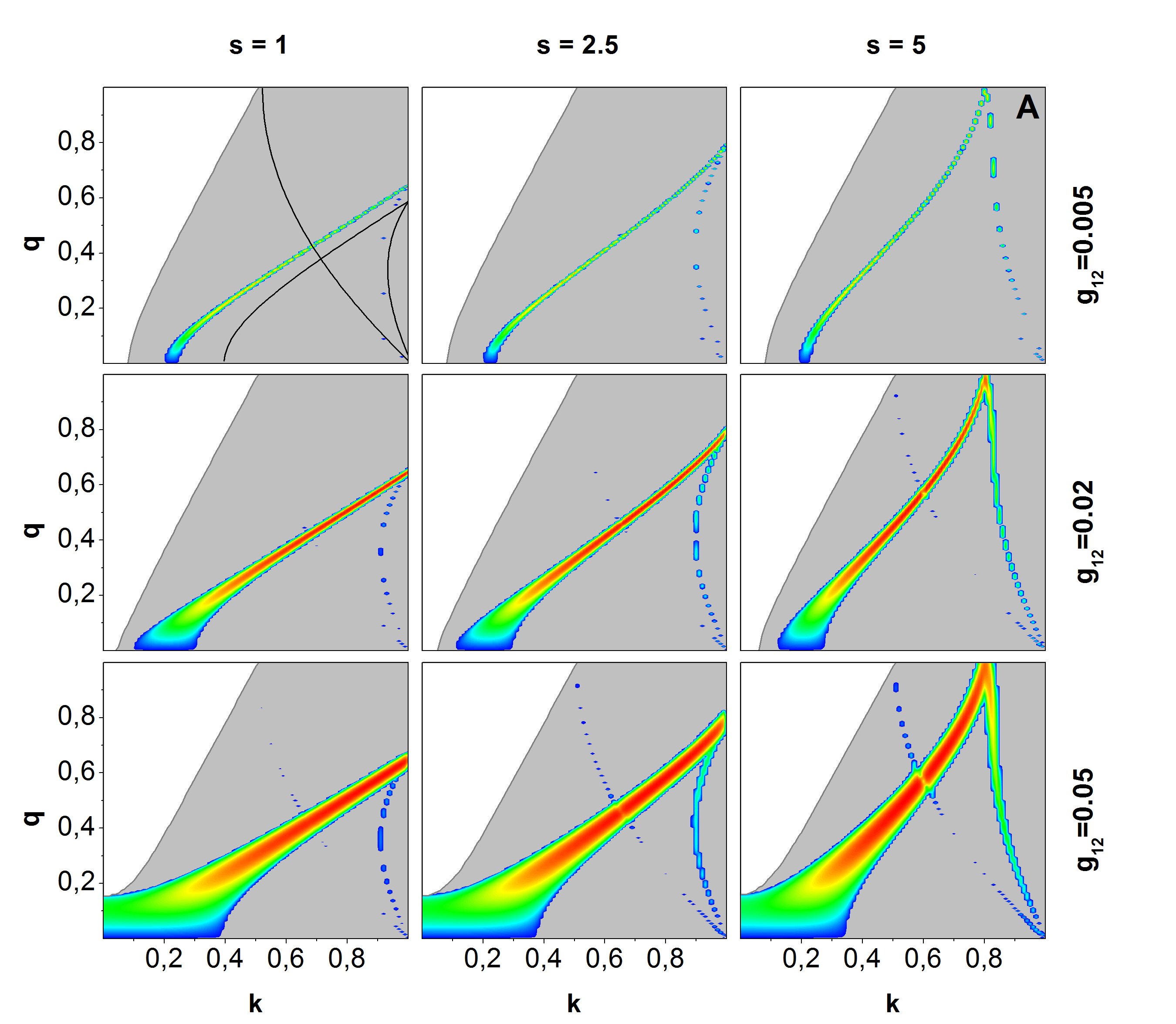}  \ \ \ \ \ \ \ \ \  \includegraphics[width=0.9\columnwidth]{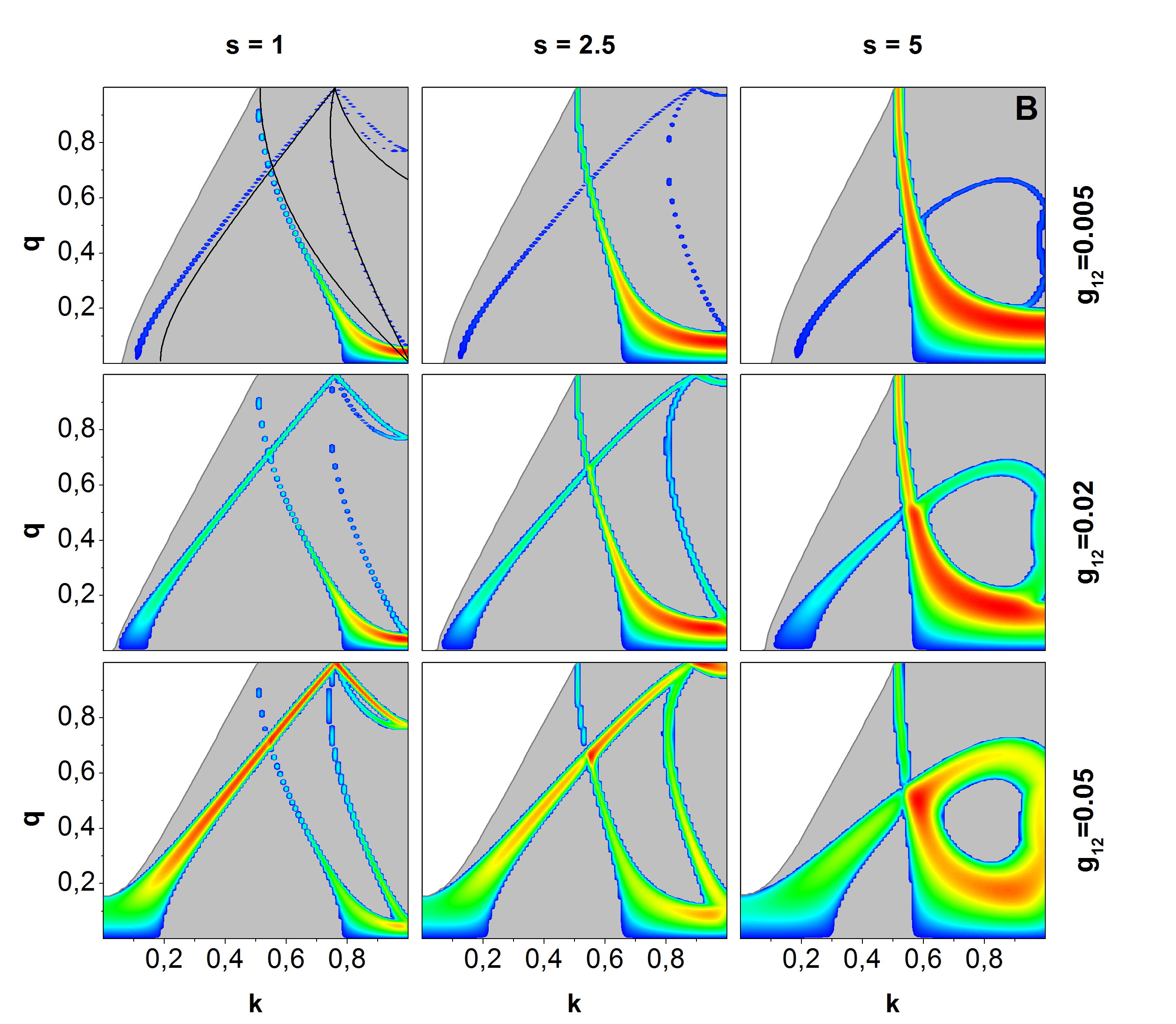}}
\caption{(Color online) Configurations \textit{A} (left) and \textit{B} (right): stability diagrams as a function of the quasimomenta $k$ and $q$ of the condensates and the excitations, for different values of $s$ and $g_{12}$. Regions in gray (color) scale indicate where the system is dynamically unstable. The gray (color) scale corresponds to the growth rate of the unstable modes ($\propto$Im($\omega_{kq}$)). Shaded area: energetic instability. Solid lines: coupling between phonon-antiphonon modes in the limit $s\rightarrow0$, $g_{12}\rightarrow0$ (see text).}
\label{fig2}
\end{figure*}

\begin{figure}
\centerline{\includegraphics[width=0.9\columnwidth]{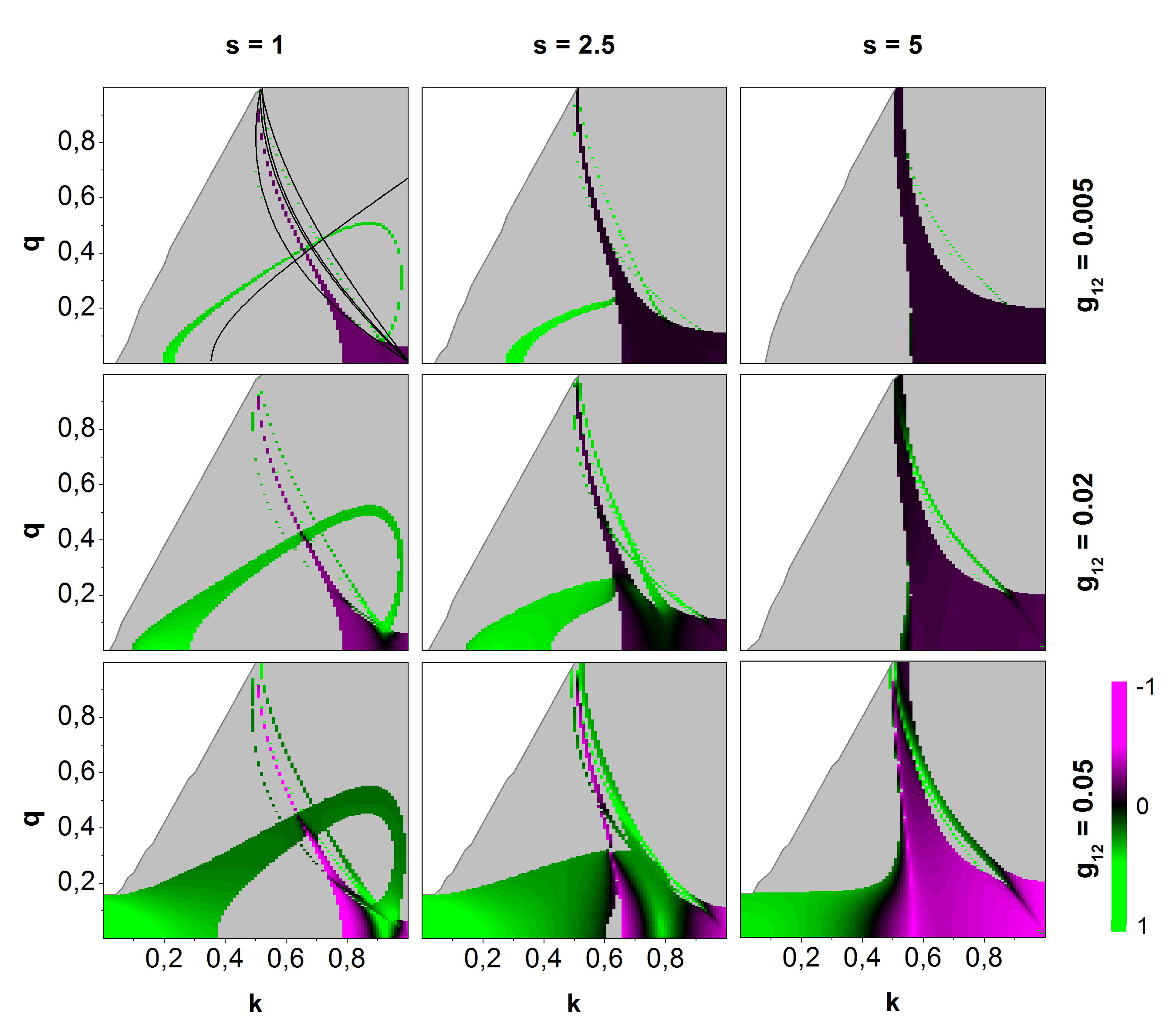}}
\caption{(Color online) Similar notations of Fig. \ref{fig2} but for configuration \textit{C}. Here the gray (color) scale corresponds to the values of $\eta(k,q)$ in the dynamically unstable regions, measuring the relative weight of each species.}
\label{fig3}
\end{figure}

The typical stability diagrams in the $k-q$ plane are shown in Fig. \ref{fig2} (for the case \textit{A} and \textit{B}) and in Fig. \ref{fig3} (case \textit{C}), for different values of the lattice intensity $s$ and interspecies interaction $g_{12}$. 
In all three cases, the energetic stability behavior is similar to that of the single species \cite{wuniu}, whereas the regions of dynamical instability show a very different and richer behavior, owing to the coupling between the two condensates, as we will explain below.
Moreover, the fact that for large enough repulsion between the two species 
the system becomes unstable also for low $k$ values (see the bottom row diagrams) is related to occurrence of a phase separation instability \cite{law,ao,hooley,ruostekoski,javanainen2} for $g_{12}>\sqrt{g_{11}g_{22}}$ 
(here the critical interaction strength is $g_{12}^{c}=0.018$).

Let us now analyze in more details the structure of the dynamical unstable regions. Since these instabilities are determined by the mixing of both components, it is instructive to know to which extent they contribute. To measure this effect here we define $V_{i}(k,q)=\int_{x}|u_{k,q}^{(i)}(x)|^{2}+|v_{k,q}^{(i)}(x)|^{2}$, and we introduce the following quantity  
\begin{equation}
\eta(k,q)=\frac{\displaystyle{V_{1}(k,q)-V_{2}(k,q)}}{
\displaystyle{V_{1}(k,q)+V_{2}(k,q)}}
\end{equation}
that is positive or negative when the major component corresponds to the BEC1 or BEC2 respectively, and close to zero when the contribution of the two BECs is almost the same.
This indicator is shown explicitly for the configuration C in Fig. \ref{fig3}, since this is the case that presents the most complex mixing behavior (due to the fact that both condensate quasimomenta are non vanishing). In particular, it emerges that the instability at low values of $k$ gets the primary contribution from BEC1, whereas that at higher $k$ is dominated by BEC2. Instead, for the configurations \textit{A} and \textit{B} of Fig. \ref{fig2} $\eta$ is mostly positive ($A$) or negative ($B$). 

\begin{figure}[t]
\centerline{\includegraphics[width=0.8\columnwidth]{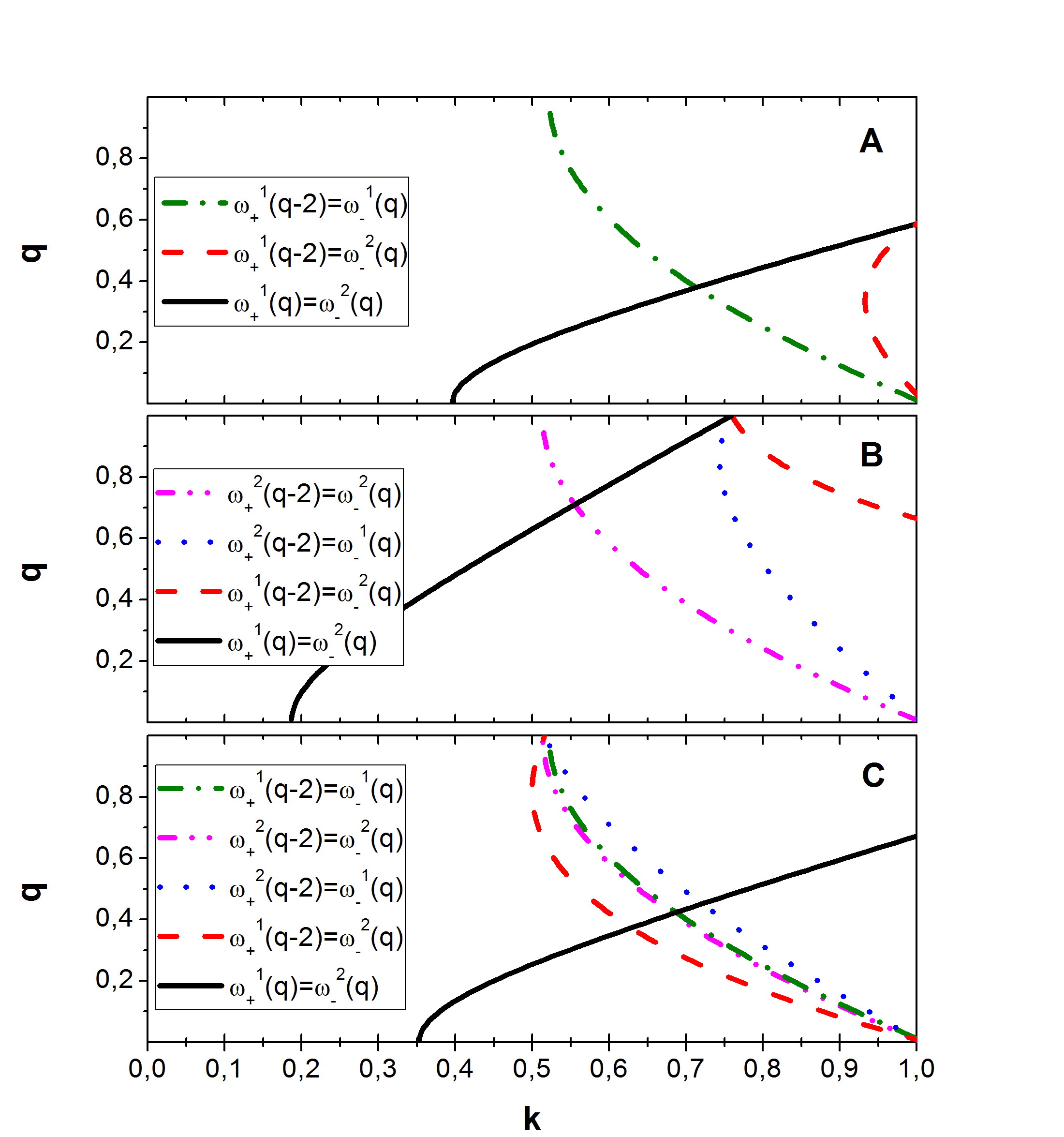}}
\caption{(Color online) Analytical solutions for the coupling of the phonon and anti-phonon modes in the limit $s\rightarrow 0$, $g_{12}\rightarrow 0$ for the three configurations A,B and C.}
\label{fig4}
\end{figure}
Further insight on the origin of the different instability regions can be gained by considering the limit $s\rightarrow 0$ and $g_{12}\rightarrow 0$. 
In this case the full matrix $\Sigma_3 M(q)$ is block diagonal, the two species are decoupled, and the  eigenvalues -- that will be indicated as 
$\omega_{\pm}^{i}(q)$  (the $\pm$ refers to the positive and negative sectors of the spectrum, $i=1,2$) -- have a simple real analytic expression in all the three configurations \cite{wuniu}.
In this limit therefore the system is dynamically stable. However, as soon as $s\neq 0$ and $g_{12}\neq 0$, a resonant coupling between phonon and anti-phonon modes may cause the appearance of an imaginary component in the spectrum, and therefore dynamical instability \footnote{Owing to the properties of the matrix $\Sigma_{3} M(q)$, complex eigenvalues can appear only in pairs, so that the onset of instability is signaled by the appearance of a pair of real degenerate eigenvalues \cite{wuniu}.}. 
Notably, in all the three cases, there are several ways to couple phonon and antiphonon modes, due to the fact that besides the usual intraspecies coupling \cite{wuniu} here we can also have interspecies resonances, namely $\omega_{+}^i(q-2)=\omega_{-}^i(q)$, 
$\omega_{+}^i(q-2)=\omega_{-}^j(q)$, and $\omega_{\pm}^i(q)=\omega_{\mp}^j(q)$.  The curves corresponding to the possible resonances lying in the range $k,q\in[0,1]$ are reported in Fig. \ref{fig4} for each of the three configurations A,B or C. 
They are also reported in the top left panels in Figs. \ref{fig2}-\ref{fig3}. Then, as we move away from the analytic limit, both by increasing the interspecies interactions $g_{12}$ or the lattice intensity $s$, the strong mixing between the different modes eventually causes a broadening and a deformation of the instability regions. 
Remarkably the seeding of most of the instabilities is the coupling of a phonon (antiphonon) mode of the species $1$ with an antiphonon (phonon) mode of the species $2$. 
This situation is completely different with respect to the case of a single condensate in an optical lattice, since the latter cannot support excitations.

Summarizing, we have investigated the stability of nonlinear Bloch waves in
a one-dimensional two component BEC system, in a configuration where one of the two condensate is modulated by an external optical lattice, effectively realizing a matter wave periodic potential for the other condensate. Such a system presents a very rich structure of the regimes of dynamical instability, originating from the resonant coupling of excitations of 
the two BECs. Different scenarios can be obtained by tuning the interspecies interaction and the relative velocity between the two condensates and/or the external optical lattice, allowing for different ways to match the resonance of phonon antiphonon branches of the two species. All these features make multicomponent ultracold atoms in optical lattices a very promising system for studying the interplay of nonlinearity and periodicity in novel regimes.

We are grateful to F. Minardi for the reading of the
manuscript and useful suggestions. We also thank S. Calamai and M. Calamai 
for helpful discussions. 
The work has been supported by EU MEIF-CT-2004-009939.

\end{document}